\documentclass[11pt]{article}

\usepackage[utf8]{inputenc}
\usepackage{newunicodechar}
\newunicodechar{ }{~}
\usepackage[backend=biber, style=numeric, sorting=none]{biblatex}
\usepackage{amsmath, amssymb}
\usepackage{graphicx}
\usepackage{hyperref}
\usepackage{geometry}
\usepackage{siunitx}
\usepackage{booktabs}
\usepackage{array}
\usepackage{url}
\usepackage{hyperref}
\usepackage{float}
\title{Advancing Resource Extraction Systems in Martian Volcanic Terrain: Rover Design, Power Consumption and Hazard Analysis}
\author{
  \textbf{Divij Gupta\textsuperscript{1}}\footnote{Email: \href{mailto:divijgupta@exospace.co.in}{divijgupta@exospace.co.in}}%
  \quad\textbf{and}\quad
  \textbf{Arkajit Aich\textsuperscript{2}}\footnote{Email: \href{mailto:arkajit.aich.kolkata@gmail.com}{arkajit.aich.kolkata@gmail.com}}\\[1ex]
  \textsuperscript{1}Welham Boys' School, Dehradun 248001, Uttarakhand, India\\
  \textsuperscript{2}Prayoga Institute of Education Research, Bengaluru 560082, Karnataka, India
}
\date{}

\begin{document}
\maketitle

\begin{abstract}
This study proposes a schematic plan for in-situ resource utilization (ISRU) in Martian volcanic terrains. The work investigated the complexity of volcanic terrains and Martian environmental hazards and suggested comprehensive engineering strategies to overcome the odds and establish a successful mining program in Martian volcanic regions. Slope stabilization methods—such as terracing and anchored drilling rigs—with terrain-adaptive rovers capable of autonomous operations on steep unstable slopes has been suggested as feasible solutions to navigate the complex geological terrains of Martian volcanoes. The mid range rover design with a mass of approximately 2.1\,t, proposed here for mining operations, incorporates a six-wheel rocker-bogie suspension, anchoring-enabled drilling arm, dust-mitigation solar arrays, and advanced sensing systems for hazard detection and navigation. A comparative analysis regarding choice of roads and rails for building transport infrastructure has also been performed. We have also looked into the energy requirement of the rover to work under extreme environmental conditions of Mars and suggested a combination of solar and nuclear power to account for the huge energy requirements of sustained operations on Mars.    
The results demonstrate that mission success in these environments depends on integrating mechanical resilience, environmental adaptability, and operational autonomy, enabling sustainable access to resources in one of Mars’ most geologically challenging settings.
\end{abstract}

\section{Introduction}

Resource extraction has long been a cornerstone of human advancement, fueling civilizations, enabling technologies, and shaping economic and geopolitical landscapes\cite{smil2017energy}. As humanity turns its gaze to planetary colonization, the imperative to utilize in-situ resources becomes paramount. Transporting all necessities from Earth is not only economically prohibitive, but logistically infeasible for long-term settlement \cite{rodriguez2021challenges}. The future of planetary habitation will depend on our ability to identify, extract, and utilize local materials for survival and infrastructure development.

Among the celestial bodies in our solar system, Mars represents the most promising candidate for near-term human colonization. Its surface conditions, diurnal cycle, and subsurface water potential offer advantages over the Moon or asteroids\cite{taylor2022marsbase}. However, sustaining human life on Mars requires far more than advanced propulsion or habitat systems. It requires a detailed understanding of Martian geology, terrain mechanics, and energy economics to make resource utilization possible. Furthermore, it needs the development of advanced technologies, substantial investment in financial and energy resources, and rigorous safety measures due to its unique environmental conditions relative to Earth.\cite{rosslowcostmars}\cite{neukart2024sustainable}. 

Due to intensive interest in Mars, there has been previous studies \cite{mit2025marsmining}\cite{rosslowcostmars} which have outlined strategic roadmap for in-situ resources extraction on Mars, detailing the development of specialized drilling tools, water extraction systems, and ISRU technologies designed for deployment by the 2030s, with projections of commercial viability following successful demonstrations. But a detailed scrutiny regarding the optimization of the energy and resource consumption in these efforts has not been done with some extraction systems requiring 24,000 kg of water making the outline commercially unprofitable. While there has been substantial scholarly work exploring the challenges and opportunities associated with human colonization of Mars, including astrogeological mining and in-situ resource utilization (ISRU), a critical gap also persists in terrain-specific analyses. Each terrain type on Mars presents itself with its unique potential and challenges which needs to be critically examined for future resource extraction programmes. In the present work, we specifically focus our attention on the volcanic terrain of Mars, which is a relatively resource-rich and mineral-rich region.

Previous research \cite{west2010potential} has demonstrated that Martian ore-forming processes—driven by hydrothermal fluids from volcanic activity, large igneous provinces, and impact craters—are analogous to terrestrial mineralization systems. This suggests that specific volcanic sites on Mars may concentrate economically valuable minerals like nickel, chromium, titanium etc all of which has widespread applications. They highlight that combining orbital spectral data with terrestrial analog models improves the identification of high-potential resource locations, a methodology critical to informed rover deployment. However, available literature did not focussed on terrain specific roadmaps for extraction of resources in volcanic regions in the most optimal way. It is crucial to note that volcanic terrains posses unique threats and require meticulous planning with regards to safe and efficient deployment of robotic systems and construction of stable infrastructure in such hazardous environments. Other vital aspects of such resource extraction programme in volcanic terrains would require a more informed approach with regards to energy requirements for slope traversal, base-to-resource distance optimization, hazard mapping, and terrain-adaptive rover design. Such a detailed analysis for extracting resources from volcanic terrains of Mars is missing and this research aims to fill that void by doing a terrain specific analysis for Martian volcanic landform and addressing all the mobility, stability, and power challenges the volcanic terrain poses. This paper is organized as follows. At first, we will present a brief overview of Martian geology with a focus on volcanic terrains, look into the resouces present in volcanic landforms and choose volcanic sites for analysis. Thereafter, we do a holistic analysis of challenges for mining volcanic terrain by incorporating hazard analysis of volcanic regions, analysing Thermal Emission Spectrometer (TES) data and exploring potential slope mining solutions for volcanic terrains. Finaly we analyze suitable Rover designs keeping in mind energy requirements and infrastructure requirements (e.g., road vs. rail transport). Overall, in this research, we will aim to provide an integrated framework for extraction of resources in one of the planet’s most geologically promising yet operationally complex environments (volcanoes) by incorporating hazard mapping, mobility analysis, and energy modeling of mining rover thereby addressing both technical constraints and mission resilience while countering Martian dust storms and environmental unpredictability.

\section{An overview of Martian Geology}

The geology of Mars exhibits a stark geological dichotomy between its southern highlands, old, heavily cratered terrains rising 1 to 4 km above datum, and the smoother, younger northern lowlands, which may once have hosted a shallow ocean. Global topographic data reveal this elevation contrast and highlight the extraordinary features of the planet's surface. Dominating the western hemisphere is the Tharsis rise, a vast igneous province approximately 4,000 km wide and 10 km high at its center, covering nearly a quarter of the Martian surface. This region is home to Olympus Mons, the tallest known volcano in the solar system, measuring 24 km in height and 550 km in diameter, dwarfing Earth’s largest volcano. These formations are widely believed to be the result of long-standing mantle plume activity beneath a stationary lithospheric plate. Adjacent to Tharsis lies Valles Marineris, an immense canyon system extending over 4,000 km and reaching depths of 10 km, with exposed stratigraphy composed largely of basaltic lava flows. Although Mars lacks active plate tectonics, its terrain hosts numerous impact basins such as Hellas Planitia which is over 2,100 km wide and 9 km deep, whose ejecta contribute significantly to southern hemisphere elevation. Despite the planet’s dramatic topographical variation, much of its geology is flat-lying, which reduces surface exposure of mineralized zones. Consequently, features like crater walls, canyon flanks, and central uplifts are prime candidates for resource exploration due to their exposed stratigraphy and increased likelihood of ore deposits.\cite{west2010potential}

\section{Martian volcanoes and their resources}

Martian volcanic terrains, particularly large igneous provinces such as Tharsis and Olympus Mons, represent highly promising sites for in-situ resource utilization. Past volcanic and hydrothermal activity has concentrated a variety of essential minerals within basaltic rocks and sand dunes.\cite{niles2024basaltic} These resources are critical for constructing habitats, producing alloys, and supporting energy systems. Additionally, hydrated minerals like gypsum and serpentine found in volcanic contexts offer potential for water extraction and environmental control.\cite{vaniman2024gypsum}\cite{emran2025serpentine} The geological richness and accessibility of these terrains make them optimal targets for resource extraction strategies aimed at supporting long-term human presence on Mars.

\subsection{Important Martian Volcanoes suitable for Resource Extraction}

The selection of Martian volcanic provinces for resource extraction is a critical step in designing terrain-adaptive infrastructure and ISRU (In-Situ Resource Utilization) strategies. Mars hosts several massive volcanic edifices, each with unique geologic and geomorphic characteristics that inform their suitability for mining applications. \textbf{Olympus Mons}, the tallest known volcano in the Solar System at over 21 km high, offers vast basaltic plains and an extensive aureole deposit field, which, despite its gentle flanks ($\sim$5$^\circ$), poses challenges related to surface dust accumulation and equipment mobility \cite{gillespie2019olympus_intrusion}. \textbf{Arsia Mons} features steep slopes, glacial deposits, and mass-wasting evidence—making it an ideal candidate for studying slope instability and regolith-ice interactions \cite{Arsiapaper}. Similarly, \textbf{Pavonis Mons}, located centrally among the Tharsis Montes, is characterized by pit craters, rift fractures, and moderately steep terrain, offering an ideal testbed for robotic anchoring and terracing solutions \cite{scott1998pavonis}. \textbf{Elysium Mons}, situated in the Elysium volcanic province, is geologically younger and exhibits well-preserved lava flow textures, valuable for analyzing excavation energy requirements and thermophysical properties \cite{elysiumpaper}. \textbf{Ascraeus Mons}, notable for its large fan-shaped lava flows and complex flow morphologies, provides insight into basaltic flow-unit mining scenarios \cite{Tharsispaper}. Lastly, \textbf{Hecates Tholus}, a smaller composite volcano, exhibits explosive-style deposits and a high-latitude location, offering unique perspectives on volatile retention and slope material cohesion under Martian conditions \cite{hauber2005hecates}. 

These six volcanoes—Olympus Mons, Arsia Mons, Pavonis Mons, Elysium Mons, Ascraeus Mons, and Hecates Tholus have been selected for comprehensive analysis in this research due to their topographic diversity, mineralogical potential, and relevance to future human exploration and ISRU operations.

\subsection{Resources in Martian Volcanoes}

Martian volcanic provinces constitute prime targets for in situ resource utilization (ISRU) owing to their composition, thermal history, and structural exposures. These terrains comprise extensive basaltic flows, large Igneous Province (LIP) architecture, and intrusive networks such as dikes and sills, which on Earth often concentrate economically relevant minerals (e.g., Fe, Cr, Ni, Ti, possibly even precious metals) through magmatic differentiation and hydrothermal activity. \cite{niles2024basaltic}\cite{mit2025marsmining}

In volcanic settings, potential mineralization includes both primary (magmatic) and secondary (hydrothermal) products. Hydrothermal alteration of volcanic rocks${-}$evidenced by silica enrichment, sulfate and hematite assemblages, and phyllosilicate alteration zones has been documented at analog sites on Earth and interpreted from Martian rover data (e.g. Columbia Hills at Gusev Crater), suggesting localized concentrations suitable for extraction.

Furthermore, analogs of magmatic-hydrothermal systems provide models for ore-forming processes on Mars; similar geologic contexts may have induced sulfide and iron-oxide deposits in altered volcanic edifices. Secondary alteration minerals—such as jarosite, hematite, opaline silica, phyllosilicates, sulfates, and clays are ubiquitous in Martian volcanic terrains and may form useful ISRU feedstocks, for example in construction or manufacturing.\cite{elysiumpaper}

\section{Challenges for Mining Volcanic Terrain}

One of the central challenges in establishing long-term operations on the volcanic terrains of Mars lies in optimizing robotic systems for resource extraction in these geologically complex terrains. These volcanic regions, while rich in subsurface water ice and economically valuable minerals like iron, nickel, and titanium, present substantial logistical and engineering difficulties. The most important factors to consider for setting up a sustainable programme of mining Martian volcanic terrains include determining the ideal placement of a base, balancing proximity to resources with safety from potential slope instabilities and environmental hazards. Furthermore, calculating the energy requirements for rovers traversing variable terrain, especially sloped volcanic flanks, is also essential to ensure sustainable operations. Terrain-informed rover design must account for mobility on basaltic rock, dust-laden surfaces, and cratered topography. Infrastructure decisions, such as the implementation of modular roads versus fixed rails, must consider dust accumulation from Martian dust storms, maintenance feasibility, and long-term adaptability in Martian weather conditions. In this section, we will take a closer look at the challenges of mining volcanic terrain of Mars and investigate the potential solutions. 

\subsection{Martian Volcanic Terrain Characteristics and Topography}
Martian volcanic terrains exhibit complex geological and topographic features which are shaped by prolonged volcanic activity, tectonic deformation, and surface weathering under Martian environmental conditions. Basaltic flows, layered pyroclastic deposits, and extensive fracturing characterize the Martian volcanic regions. Overall, the volcanic regions combine gently sloping shield volcano flanks (typical gradients $ \sim 5^{\circ} {-}7^{\circ}$) with rugged terrains featuring fissures, collapse pits, and ash-laden soils. 

Among Martian volcanic terrains, Olympus Mons, the tallest volcano in the Solar System, is one of the most studied. However, Olympus Mons is not the best representative of slope instability challenges of Martian volcanic terrain. Instead, Arsia Mons and Pavonis Mons in the Tharsis region present more immediate concerns for terrain stability due to their steep flanks, layered pyroclastic deposits, and possible presence of ice-rich regolith.\cite{https://doi.org/10.1029/2022JE007467}. These volcanoes in particular feature complex geomorphology including flank terraces, pit craters, and landslide scars, indicating active or past mass-wasting events. The surface composition is largely basaltic, with grain sizes ranging from dust to volcanic breccia, depending on the degree of erosion and past eruptive style. The topography is highly uneven, with slopes seldom exceeding $\ang{15}{-} \ang{25}$ along caldera rims and flank scarps \cite{https://doi.org/10.1029/2022JE007467}. Slope stability in these regions is further compromised by the presence of porous volcanic tuff, low-cohesion material, and potential subsurface ice, which may lead to basal lubrication and failure under mechanical stress. 

Due to the complexity of Martian volcanic landforms, accurately characterizing such terrains is foundational for mining feasibility studies. 
 
\subsection{Slope Stability Analysis}

\begin{table}[H]
\centering
\renewcommand{\arraystretch}{1.2}
\begin{tabular}{|p{3.5cm}|p{2.5cm}|p{2.5cm}|}
\hline
\textbf{Volcano} & \textbf{Latitude} & \textbf{Longitude} \\
\hline

\multicolumn{3}{|c|}{\textbf{MOLA Tile: \SI{0}{\degree}--\SI{44}{\degree}N, \SI{180}{\degree}E--\SI{270}{\degree}E, Topography File: \texttt{megt44n180hb.lbl}}} \\
\hline
Olympus Mons     & \SI{\sim 18.6}{\degree}N & \SI{\sim 226}{\degree}E \\
Pavonis Mons     & \SI{\sim 0.8}{\degree}N  & \SI{\sim 247}{\degree}E \\
Ascraeus Mons    & \SI{\sim 11.9}{\degree}N & \SI{\sim 255.5}{\degree}E \\
\hline

\multicolumn{3}{|c|}{\textbf{MOLA Tile: \SI{0}{\degree}--\SI{44}{\degree}S, \SI{180}{\degree}E--\SI{270}{\degree}E, Topography File: \texttt{megt00n180hb.lbl}}} \\
\hline
Arsia Mons       & \SI{\sim 8.3}{\degree}S  & \SI{\sim 239}{\degree}E \\
\hline

\multicolumn{3}{|c|}{\textbf{MOLA Tile: \SI{0}{\degree}--\SI{44}{\degree}N, \SI{90}{\degree}E--\SI{180}{\degree}E, Topography File: \texttt{megt44n090hb.lbl}}} \\
\hline
Elysium Mons     & \SI{\sim 25}{\degree}N   & \SI{\sim 147}{\degree}E \\
Hecates Tholus   & \SI{\sim 32.9}{\degree}N & \SI{\sim 150}{\degree}E \\
\hline
\end{tabular}
\caption{Grouping of Martian volcanoes by corresponding MOLA tiles and topography files.}
\label{tab:volcano_mola_tiles}
\end{table}

To ensure the safety and security of mining operations in the volcanic terrains of Mars, it is of utmost importance to perform a slope stability analysis of the complex topography of volcanic landforms. 

To evaluate the stability of Martian volcanic terrain, we utilized topographic data from the Mars Orbiter Laser Altimeter (MOLA), a key instrument aboard NASA’s Mars Global Surveyor. The MOLA Experiment Gridded (MEGT) dataset\cite{mola_megdr} provides high-resolution digital elevation models (DEMs) with vertical accuracy on the order of meters and horizontal resolutions of up to 463 meters per pixel, making it highly suitable for terrain gradient analysis at regional scales. For the purpose of slope stability analysis, we focused on the high-resolution MOLA grids covering Olympus Mons, Pavonis Mons, and Ascraeus Mons, which were accessed in '.img' format and geo-referenced using their corresponding '.lbl' metadata descriptors.

The slope stability assessment is grounded on a quantitative interpretation of terrain parameters extracted from the DEM, as summarized in the Table 1. These metrics provide critical input for evaluating slope-induced hazards and for identifying potential sites for future base or infrastructure placement.

Python was employed as the primary computational platform for this analysis, using scientific libraries such as GDAL for geospatial raster processing, NumPy for numerical computation, and Matplotlib for data visualization. Slope values were calculated by applying a central difference gradient method to the elevation arrays derived from the MOLA DEMs, enabling the generation of slope maps that visually represent variations in terrain inclination across volcanic flanks and caldera rims.

The slope stability analysis includes the \textit{Factor of Safety (FoS)}, which is a critical geotechnical parameter that quantifies the stability of a slope by comparing resisting forces to driving forces acting on a potential failure plane. In the context of Martian terrain, where reduced gravity and regolith variability pose significant challenges, FoS provides a first-order approximation of slope reliability. A slope is considered stable if the FoS $>$ 1.0 and unstable if FOS $<$ 1.0. For this study, the infinite slope model was applied to determine FoS values using terrain data derived from MOLA digital elevation models. Mathematically, FOS is defined as \cite{Duncan2014Soil},

\begin{equation}
\text{FoS} = \frac{c + \gamma z \cos^2\theta \tan\phi}{\gamma z \sin\theta \cos\theta}
\label{FoS}
\end{equation}

\noindent where $c$ is the cohesion of the regolith (Pa). In rover and simulation studies \cite{AngleFriction},  c is estimated to lie in the range of 1–5 kPa. However, we have set it at 2000 Pa, reflecting the weakly cemented nature of basaltic Martian soil. In equation \eqref{FoS}, $\gamma$ is the unit weight (N/m$^3$), calculated as the product of density of martian regolith and gravity; $z$ is the depth of the failure plane (m), assumed to be \SI{1}{\metre} for unit thickness; $\theta$ is the slope angle (degrees); and $\phi$ is the internal friction angle (degrees), assumed to be \SI{30}{\degree} which is a typical value for dry, granular, angular basaltic materials observed on planetary bodies and falls within the range used in terrestrial slope analyses \cite{AngleFriction}. This formulation assumes a planar, shallow failure surface parallel to the slope, consistent with dry, granular Martian regolith conditions. A similar approach has been adopted by Fenton et al. \cite{fentonlandslide}, who utilized remote sensing datasets to assess landslide hazards and label ‘danger zones’ which are prone to landslides. Their results underscore the applicability of FoS mapping for planetary geotechnical risk assessment.

To add another layer of analysis, we considered the superimposition of aspect maps with topographic contour lines, which serves as a robust methodological approach to characterize planetary surface morphology and terrain orientation. Aspect, defined as the azimuthal direction of the steepest slope at a given point, plays a pivotal role in modulating surface energy balance, insolation exposure, aeolian erosion patterns, and microclimatic variability—factors of critical significance for long-term robotic or crewed installations on extraterrestrial bodies such as Mars. Topographic contours, by delineating lines of constant elevation, facilitate the visualization of slope gradients, elevation discontinuities, and geomorphological features. The combined representation of these two spatial variables enhances the interpretive capacity of geomorphic assessments by correlating slope directionality with local relief features. In the present analysis, aspect was computed via the arctangent of orthogonal DEM gradient components, while contours were rendered through evenly spaced elevation thresholds, thus producing a composite spatial framework. This integrative visualization aids in the identification of mechanically stable and environmentally shielded zones, informing hazard mitigation strategies and optimal site selection protocols in planetary geotechnical evaluations.

\begin{figure}[H]
    \centering
    \includegraphics[width=\linewidth]{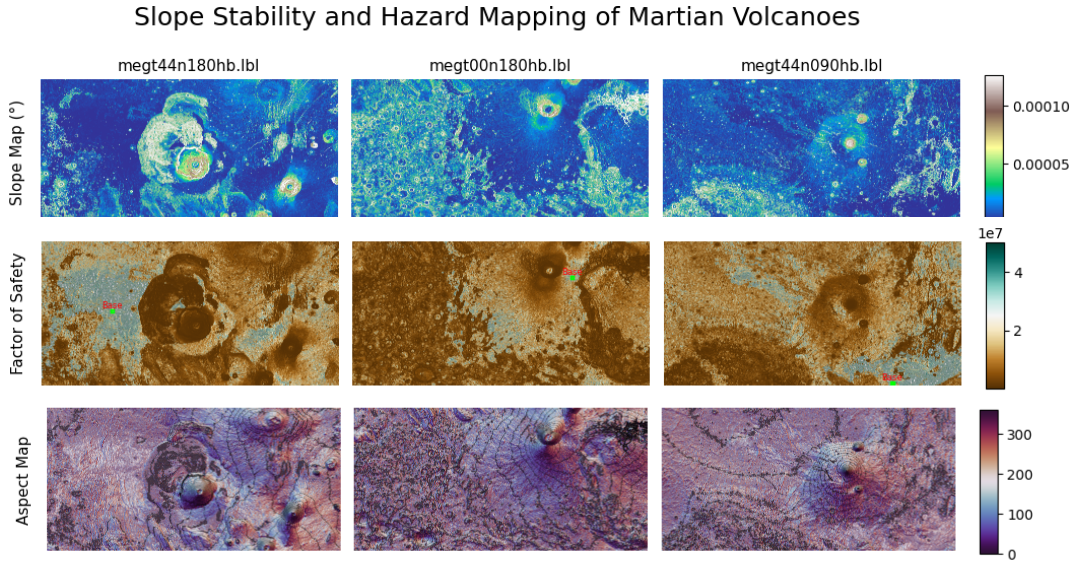}
    \caption{Hazard mapping of the volcanic terrain on Mars.}
    \label{fig:hazard_map}
\end{figure}

In Figure 1, the first column represents the dataset megt44n180hb.lbl, which contains the topographical data for the martian geographical region \SI{0}{\degree}--\SI{44}{\degree}N, \SI{180}{\degree}E--\SI{270}{\degree}E and covers volcanic terrain of Olympus Mons, Pavonis Mons and Ascraeus Mons; second column represents the dataset megt00n180hb.lbl, which contains the topographical data for the martian geographical region \SI{0}{\degree}--\SI{44}{\degree}S, \SI{180}{\degree}E--\SI{270}{\degree}E covering Arsia Mons volcanic terrain; and the third column represents the dataset megt44n090hb.lbl, which contains the topographical data for the martian geographical region \SI{0}{\degree}--\SI{44}{\degree}N, \SI{90}{\degree}E--\SI{180}{\degree}E and encompases Elysium Mons and Hecates Tholus volcanic regions. Top row represents Slope Map in degrees; Middle row represents Factor of Safety (FoS), indicating mechanical stability and the appropriate location for Martian Base; and Bottom row represents Aspect maps overlaid with elevation contours to infer orientation, volcanic flow and topographic variability.

From the slope map, it is evident that overall, the terrain exhibits very low slope angles (mostly $< 2^\circ$). This is expected due to the large spatial resolution and gradual elevation changes of Martian volcanic shields. Olympus Mons shows slightly elevated slopes near caldera rims and flank scarps, consistent with its steep summit geometry. Arsia Mons and Elysium Mons exhibit more uniform and subdued slope profiles.

In the middle row, the FOS is represented. Here the slope stability is quantified using an infinite slope model assuming dry, cohesion-driven regolith conditions. The FOS values are exceptionally high across all three regions (typically $>10^7$), reflecting low slope angles. Localized dips in FoS correspond to minor topographic features like caldera walls or erosional troughs. Although large-scale failure is unlikely in most Martian volcanic terrains, estimation of FOS helps exclude subtle unstable zones and prioritize sites for sustained surface operations or subsurface construction. 

The calculation of FOS utilized the MOLA MEGT gridded dataset which provides topographic elevation with a resolution of 128 pixels per degree. The mean radius of Mars \( R = 3,396,190\ \text{m} \) is also used in the calculation as a parameter. The spatial resolution in meters per pixel is calculated as:

\[
\text{Meters per pixel} = \frac{2\pi R}{360 \times N}
\]
\[
\text{Meters per pixel} = \frac{2\pi \times 3,396,190}{360 \times 128} \approx 463\ \text{m/pixel}
\]

A moving average kernel of size \(40 \times 40\) pixels was used for assessing regional stability and identifying optimal base locations. At 463 meters per pixel, this corresponds to an approximate analysis window of:

\[
\text{Kernel size} = 40 \times 463\ \text{m} = 18.52\ \text{km}
\]

This approach of estimating FOS allows assessment of stability over mesoscale regions relevant to human or robotic surface activity planning. A smoothing kernel of approximately $40 \times 40$ pixels (covering a $\sim18.5 \times 18.5$ km region) was applied using a uniform filter to estimate averaged stability over potential base site footprints. Regions with the highest local FoS averages were outlined in green boxes as candidate zones for future surface installations.

The third row combines slope aspect maps—showing the direction of maximum slope—with elevation contours to produce a dual-layered visualization. This composite representation enhances the ability to interpret terrain orientation, relief, and drainage-like structures. Olympus Mons exhibits radial symmetry in aspect values, suggesting homogeneous flank slopes, while Elysium Mons shows more disrupted patterns likely due to erosional or secondary volcanic processes. The areas colored in red ($\geq 180$ on the color bar) are classified as leeward slopes or thermally sheltered sites which are safe from any volcanic lava flow. On the contrary, areas in blue ($\leq 80$ on the color bar) are prone to lava flow in case of any volcanic eruption. However, these areas are likely to remain safe for at least a couple of centuries due to the dormancy of the volcanoes. Combined aspect-contour maps are also essential for orientation-dependent decisions, such as aligning solar panels, or anticipating dust transport dynamics. This layer adds terrain intelligence that cannot be gleaned from elevation data alone.

This visualization framework enables integrated terrain analysis for Martian surface planning. By combining topographic, geotechnical, and directional indicators, the framework facilitates robust site selection for future landing missions, habitat placement, or ISRU operations. The methodology is adaptable to other Martian locales and planetary bodies with sufficient topographic datasets.

\subsection{TES Data Analysis}

In the previous section, we performed a slope stability analysis on the most important volcanoes of Mars. To have further insights on the volcanic terrains, we processed Thermal Emission Spectrometer (TES) data to investigate spatial and spectral variations in surface emissivity and thermal properties of Martian volcanic regions. However, due to scarcity of such data, we have limited our TES data analysis to only one volcanic region - Olympus Mons.   

The raw TES dataset, provided in tabular form, was first parsed to extract geospatial coordinates (latitude and longitude) and calibrated spectral radiance values across multiple bands. For the purpose of this analysis, the first four spectral bands (Band 1 to Band 4) were selected, as they represent key wavelengths in the mid-infrared region that are sensitive to surface mineralogy and thermal inertia.

\begin{figure}[H]
    \centering
    \includegraphics[width=\linewidth]{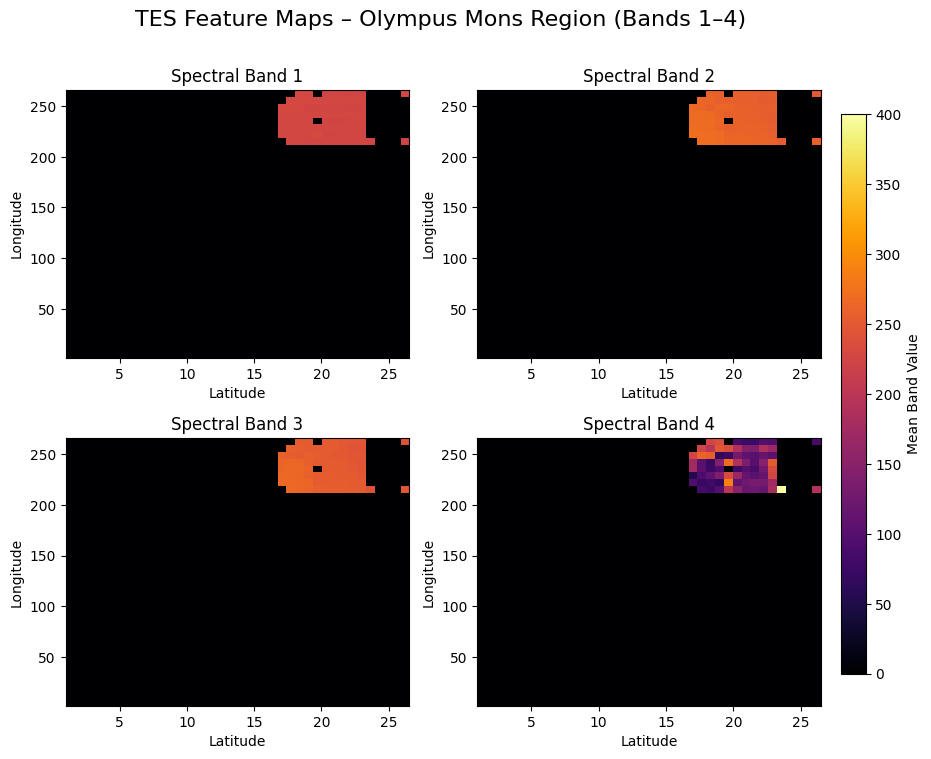}
    \caption{TES Feature Maps for Olympus Mons Region (Bands 1–4)}
    \label{fig:tesanalysis}
\end{figure}

To spatially aggregate the spectral data, a two-dimensional binning procedure was implemented. Latitude and longitude values were discretized into uniform grid cells, and the mean spectral radiance for each band was computed within each cell. This approach effectively reduces random noise while preserving the spatial distribution of radiometric features. Only regions with valid TES observations were retained; empty bins were masked out to avoid bias in the spatial representation.

The resulting mean spectral maps (Fig.~\ref{fig:tesanalysis}) reveal that the majority of spectral activity is concentrated within a compact, high-intensity zone in the northwestern quadrant of the study area. This region exhibits elevated radiance values across all four bands, with particularly strong responses in Bands 1, 2 and 3 suggesting the presence of thermally distinct surface materials or localized compositional heterogeneity. Band 4 displays a more heterogeneous spatial pattern, possibly indicating a mix of mineralogical endmembers with distinct emissivity characteristics.

The persistent spatial coherence of the high-intensity region across all four bands suggests a geological or thermophysical anomaly, such as exposed rock outcrops, pyroclastic deposits, or recent lava flows. This is consistent with the geological features of Olympus Mons. Future work will involve spectral unmixing and comparison with laboratory emissivity spectra to constrain the mineralogical composition of these regions and to infer their geologic origin. The analysis can be extended to other volcano candidates in follow-up works.

\subsection{Mining Techniques and Integrated Solutions}
Techniques for ensuring operational stability, resource access, and robotic autonomy must be tailored to account for Mars’ reduced gravity, abrasive dust, and the mechanical behavior of basaltic soils. In this section, we highlight the mining strategies that can be implemented for extracting resources from the volcanic terrains of Mars.

\textbf{Terracing}

Terracing involves the construction of horizontal steps or platforms cut into sloped terrain, which can serve as stable bases for the deployment of equipment and material extraction. This method reduces the angle of repose locally and helps minimize the risk of landslides or equipment slippage. On Earth, terracing is commonly used in surface mining operations in hilly regions to improve the safety of workers and machinery. Its adaptation for Mars has been explored in conceptual designs for regolith excavation systems, where stability and redundancy are prioritized for autonomous excavation in volatile slopes. \cite{mueller2022regolith_excavation}

\textbf{Anchored Drilling Rigs}

Anchored drilling systems offer mechanical stability by distributing loads and resisting displacements caused by drilling vibrations or gravity-induced shifts. In Martian applications, anchoring solutions such as self-drilling anchors or grouted rods can be embedded into regolith to provide localized stabilization for robotic or semi-autonomous rigs.\cite{inproceedings} These systems have shown effectiveness in stabilizing terrestrial slopes and are being considered for extra-planetary adaptations in extra-terrestrial construction engineering.\cite{kampas2020extraterrestrial_anchors} 

\textbf{Robotic Assistance}

Robots capable of operating on steep or unstable slopes can enhance mining operations by performing repetitive or hazardous tasks remotely. Tethered or legged robotic platforms have been proposed for inclined regolith collection, micro-drilling, and payload transport. Their deployment reduces the risk to astronauts and allows for persistent operation in otherwise inaccessible zones.\cite{kolvenbach2021spacebok} Studies demonstrate that climbing robots equipped with adhesion or grappling mechanisms can operate in simulated Martian environments, providing foundational technology for high-slope exploration.\cite{kolvenbach2021spacebok}

\textbf{Remote Monitoring}

The integration of remote sensing technologies—particularly ground-penetrating radar (GPR), LiDAR, and stereo photogrammetry—enables real-time assessment of terrain deformation and early detection of slope instabilities. These systems provide crucial feedback for operational planning and risk mitigation. Techniques such as Synthetic Aperture Radar Interferometry (InSAR) have been proposed for Martian analog studies to monitor landslide-prone regions in volcanic provinces like Arsia Mons.\cite{shi2020insar_landslide_review}

\textbf{Design Considerations for Terrain-Adaptive Rovers}

Mining platforms and exploratory vehicles must incorporate specialized mobility and structural design features to function effectively on sloped Martian surfaces. Rocker-bogie suspension systems provide mechanical flexibility for obstacle negotiation, while traction can be enhanced through wheels fitted with grousers or cleats for improved grip on loose basaltic soil. Additionally, a low center of gravity improves stability, reducing the likelihood of tipping. Autonomy is also critical; terrain-adaptive algorithms for obstacle avoidance and path optimization are necessary due to communication delays. NASA's All-Terrain Hex-Legged Extra-Terrestrial Explorer (ATHLETE) rover embodies many of these principles, serving as a prototype for large-scale, modular surface operations.\cite{wilcox2011athlete}

\section{Analyzing Efficiency for Transport Infrastructure}

In the previous sections, we looked into the challenges for mining volcanic regions of Mars and explored the possible solutions. Apart from the challenges of Martian terrains, the transportation of resources is another key aspect of a successful mining program. In this section, we pay our attention to the transport infrastructure and the challenges associated with it.

\subsection{Roads vs. Rails for Movement of Martian Rover}
Infrastructure for rover movement on Mars requires a strategic decision between roads and rails, each offering unique advantages and challenges. Roads, comprised of compacted regolith or sulfur concrete, can be laid quickly using in-situ materials, offering flexibility in routing and adaptability as mission needs evolve.\cite{LIU20221} However, their loose surface is prone to dust accumulation and surface degradation, necessitating frequent maintenance (e.g., grading, regolith compaction).\cite{rechyonarticle2019} Conversely, rails, especially steel or maglev systems—provide more energy, efficient transport due to lower rolling resistance, but their construction requires precise bed preparation and greater initial resource investment.\cite{su16073041} Rails are also less adaptable to changing routes and may still be affected by dust and thermal expansion issues in the Martian environment. 

From an engineering perspective, any roadbed robust enough to support rail infrastructure would inherently possess the structural integrity necessary to accommodate wheeled vehicles as well. While rail systems, characterized by steel wheels on steel tracks, offer significantly lower rolling resistance compared to traditional rubber-tired vehicles,\cite{su16073041} this advantage is partially offset on Mars by the impracticality of using rubber due to its transition to a brittle, glass-like state at low Martian temperatures. Consequently, the materials and design configurations for Martian wheeled mobility systems remain an active area of research and development.

\subsection{Impact of Martian Dust on Roads and Rails and Prevention Strategies}

Based on data from regional dust storm occurrences in the mid-latitude region, the average rate derived from Martian climate models is approximately 0.15 storms per sol.\cite{TexturedDustStormActivityinNortheastAmazonisSouthwestArcadiaMarsPhenomenologyandDynamicalInterpretation} This aligns with observations that local storms (10\textsuperscript{3}–10\textsuperscript{6} km\textsuperscript{2} in size) occur frequently each Martian year during southern hemisphere spring–summer, which is about one event per six to seven sols on average.\cite{TexturedDustStormActivityinNortheastAmazonisSouthwestArcadiaMarsPhenomenologyandDynamicalInterpretation}

For roads, we used a baseline efficiency loss of 0.2\% per sol, supported by Martian rover solar panel degradation studies.\cite{2021P&SS..20705337L}

For rails, the calculation for baseline efficiency loss is a little complex. Analysis of Mars Exploration Rover (MER) calibration targets reports dust accumulation at approximately $0.004$ optical depth per sol.\cite{LORENZ2021105337} Comparing this with solar array degradation rates ($\sim0.2\%$/sol) suggests comparable ground$-$level deposition. Research on dust buildup over the UV sensor and Mastcam camera deck shows that horizontal deck surfaces accumulate dust more rapidly; elevated or angled components accumulate significantly less ($40$$-$$70\%$ of deck rate).\cite{LorenzRalph} Based on these findings, we approximate the dust deposition rates for elevated rail structures at $0.08\%$ efficiency loss per sol, which is approximately $40\%$ of the rate observed for ground$-$level roadways. 

\begin{figure}[H]
    \centering
    \includegraphics[width=0.8\textwidth]{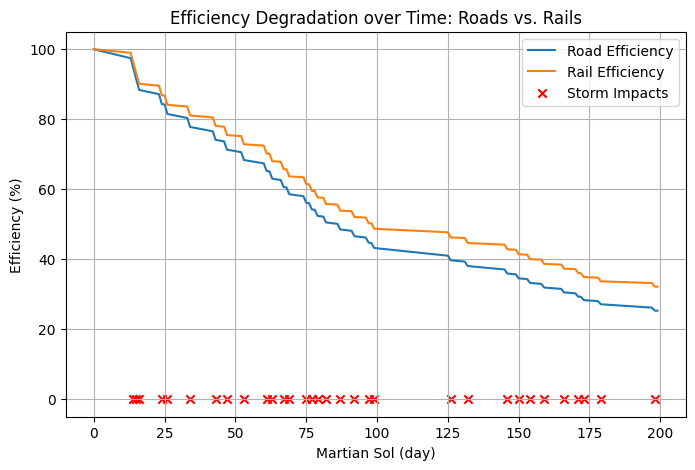}
    \caption{Efficiency degradation of roads and rails under daily wear and Martian dust storm events. Red crosses mark the days when storms occur according to probabilistic assumptions.}
    \label{fig:roadvsrail}
\end{figure}

Using Python’s NumPy and Matplotlib libraries, Figure 1 illustrates a line graph comparing the efficiency degradation of roads and rails over time on Mars. The x-axis represents Martian sols (days), while the y-axis indicates infrastructure efficiency as a percentage. The blue and orange curves correspond to the performance of roads and rails, respectively. While both systems exhibit similar initial degradation due to daily wear and dust impacts, rails maintain higher efficiency over extended durations. This trend suggests that mission duration plays a critical role in infrastructure planning. Shorter missions may favor roads for their lower construction cost and flexibility, whereas longer missions would benefit from the durability and long-term performance of rail systems.

Infrastructure and dust pose major operational challenges for Martian transit systems. Dust accumulation can quickly cover transport surfaces, reducing traction and obscuring mechanical components, especially on ground-level roads where settlement is high. Moreover, electrostatic adhesion exacerbates this issue: fine dust particles, often triboelectrically charged, cling stubbornly to metal and polymer surfaces upon contact. 

To mitigate these effects, several strategies can be adopted. Surface coatings, such as anti-static or electrodynamic dust shields, can significantly reduce dust adhesion by neutralizing surface charges.\cite{harrison2016applications}Regular maintenance, including mechanical brushing or compressed air systems, helps remove dust buildup and restore operational capability.\cite{cho2025dust} Additionally, design adaptations, such as elevating tracks or enclosing pathways, can limit dust deposition by exploiting wind shear layers close to the ground. Implementing redundant routes further enhances mission resilience by providing alternate paths when primary routes are compromised by storms or terrain hazards. By combining strategic route planning, informed by historical storm patterns and terrain analysis\cite{whetsel2025utilizing}, with dynamic on-board routing, rovers can proactively avoid compromised zones and optimize their pathways in real-time. This integrated approach ensures continuity of operations during dust events and distributes mission risk across multiple infrastructure elements.

\section{Rover design}

With the transport infrastructure strategies put forward in last section, we now proceed with the design of the rover which is the most important aspect of Martian resource mining at volcabbnic regions. 

\subsection{Features of the Rover}
The proposed Martian slope-mining rover has been designed to operate in geologically complex and hazardous environments such as volcanic flanks, crater rims, and regolith-covered slopes. Its features reflect a synthesis of mechanical stability, environmental adaptability, and autonomous functionality.

To navigate steep inclines and uneven terrain, the rover employs a six-wheel rocker-bogie suspension system. This configuration minimizes chassis tilt and ensures continuous ground contact, enhancing traction and stability. The wheels are outfitted with grousers to increase grip on fine regolith and loosely consolidated volcanic tuffs.

At the front of the chassis is a multi-jointed robotic drilling arm capable of handling core extraction and geotechnical analysis. The arm integrates a self-drilling anchoring mechanism inspired by terrestrial grouted anchor systems, enabling stable operation on inclined surfaces without slippage. This allows safe drilling in slope-prone zones where unanchored rigs would be at risk of tipping or shifting.

The rover is designed to operate on artificially terraced platforms created as part of the slope-mining protocol. These terraces reduce shear stress on the surrounding soil and provide level workspaces. The rover’s footprint and wheelbase are optimized to remain stable within the bounds of each terrace step, ensuring secure positioning during high-force operations like drilling or robotic manipulation.

Martian dust poses a persistent threat to surface operations. To mitigate this, the rover features solar panels embedded with programmable piezoelectric micro-actuators. These actuators generate localized micro-vibrations that dislodge dust particles from panel surfaces and joints without requiring moving parts or active airflow, thereby improving energy efficiency and minimizing mechanical failure points.

A telescopic mast equipped with remote sensing instruments, including LiDAR and radar modules, enables continuous assessment of surrounding topography and slope stability. These systems provide data critical to hazard avoidance, route planning, and terrain modeling, especially in unstable environments where manual intervention is impossible.

To further improve slope resilience, the rover maintains a low center of gravity and a compact horizontal profile. This structural design reduces the risk of toppling on steep gradients and facilitates anchoring on variable slope geometries. Reinforced wheel hubs and articulated linkages improve load distribution across inclined surfaces.

The rover’s contact mechanics and movement algorithms are tuned to operate effectively on basaltic regolith with varying grain sizes and cohesion levels. Its wheel and suspension design have been stress-tested in Martian analog environments using finite element modeling and soil interaction simulations.

The target mass for the slope-mining rover is driven by mission capability requirements rather than arbitrary sizing. Key drivers are: (1) mobility and traction for steep, regolith-covered volcanic slopes (necessitating a robust suspension, large wheels and high-torque actuators); (2) a multi-jointed manipulator and coring capability with an integrated anchoring system to permit stable, high-force drilling operations on inclined terraces; (3) a power system sized for continuous operations in dust-prone environments (choice between large solar arrays with batteries or an RTG + battery buffer); (4) a complement of remote sensing instruments for continuous terrain assessment (LiDAR, small GPR, spectrometers); and (5) thermal, avionics and communications hardware sized for long endurance and redundancy.

\begin{table}[H]
\centering
\label{tab:component_breakdown}
\renewcommand{\arraystretch}{1.2}
\begin{tabular}{|l|r|r|}
\hline
\textbf{Subsystem} & \textbf{Mass range (kg)} & \textbf{Chosen mass (kg)} \\
\hline
Mobility \& suspension (6-wheel rocker-bogie + reinforced hubs) & 120--450 & 350 \\
Chassis, frame and reinforcement (anchoring interface) & 150--600 & 400 \\
Robotic arm + turret (multi-joint) & 60--300 & 200 \\
Drill, coring and anchor systems & 10--120 & 60 \\
Primary power unit (RTG) & 43--45 & 45 \\
Battery + power electronics + dust-cleaning actuators & 80--300 & 200 \\
Instruments (LiDAR, small radar, spectrometers, cameras) & 50--250 & 150 \\
Wheels, motors and drive hardware (separate) & 60--300 & 200 \\
Avionics, thermal control and communications & 40--300 & 150 \\
\hline
\textbf{Subtotal} &  & \textbf{1,755} \\
Contingency (20\%) &  & \textbf{351} \\
\textbf{Total (design mass)} &  & \textbf{2,106 kg} \\
\hline
\end{tabular}
\caption{Component mass breakdown used to derive the mid-range rover mass.}
\end{table}
Mass estimates for each functional subsystem were chosen from realistic engineering analogues (flight and prototype rovers, lightweight drill designs and existing power-source masses) and scaled upward to account for slope-optimized structural reinforcement and anchoring hardware. A conservative allocation of margin (20\%) is applied to the aggregated subsystem mass to cover harnessing, fittings, localized reinforcement, and other non-modeled items (fasteners, connectors, thermal blankets). The final chosen design mass is therefore the sum of the selected component masses plus the contingency margin:
\[
M_{\mathrm{sub}}=\sum_i m_i,\qquad
M_{\mathrm{cont}} = 0.20\cdot M_{\mathrm{sub}},\qquad
M_{\mathrm{total}} = M_{\mathrm{sub}} + M_{\mathrm{cont}}.
\]
Using the component values in Table~\ref{tab:component_breakdown} results in \(M_{\mathrm{sub}}=1{,}755\ \mathrm{kg}\), \(M_{\mathrm{cont}}=351\ \mathrm{kg}\), and \(M_{\mathrm{total}}=2{,}106\ \mathrm{kg}\) \(\approx2.1 t\). This mid-range mass provides sufficient margin to support moderate ISRU interfaces (drill, anchoring, and sample handling), an RTG or robust battery pack, and hardened mobility systems appropriate for volcanic flank operations.
\begin{table}[H]
\centering
\label{tab:scenario_summary}
\renewcommand{\arraystretch}{1.2}
\begin{tabular}{|>{\centering\arraybackslash}p{4.5cm}|>{\centering\arraybackslash}p{3cm}|>{\centering\arraybackslash}p{3cm}|}
\hline
\textbf{Scenario} & \textbf{Mass range (kg)} & \textbf{Representative design mass (kg)} \\
\hline
Light demonstrator (solar, small arm, limited ISRU) & 770--1,530 & 1,000 \\
Mid-range (this study; robust drill + RTG/battery buffer) & 1,650--2,500 & \textbf{2,106} \\
Heavy ISRU (smelter/processors, extended life) & 1,650--4,100 & 2,500--3,000 \\
\hline
\end{tabular}
\caption{Scenario mass ranges and recommended design points for trade-space comparison.}
\end{table}

The mid-range design having a mass of approximately 2.1\,t balances capability and launch/cost constraints: it supports a robust manipulator and anchored drilling while retaining the potential to include a single MMRTG for continuous electrical power and moderate battery buffering.

\begin{figure}[H]
    \centering
    \includegraphics[width=0.8\textwidth]{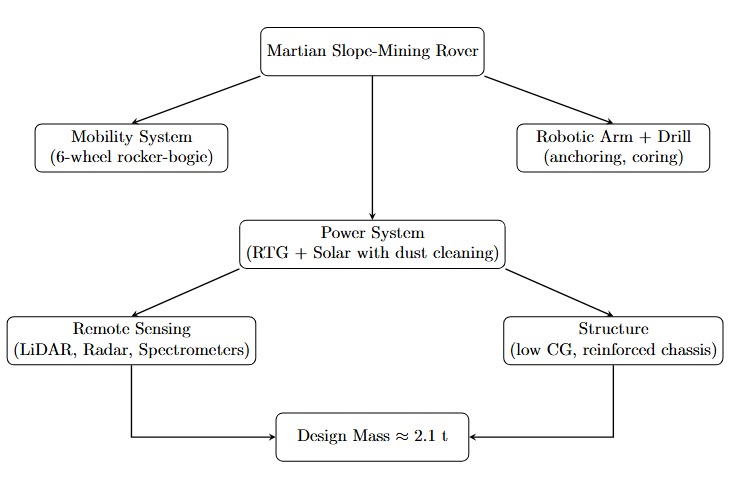}
    \caption{Schematic diagram of rover highlighting important functionalities}
    \label{fig:Schematic}
\end{figure}

\begin{figure}[H]
    \centering
    \includegraphics[width=0.8\textwidth]{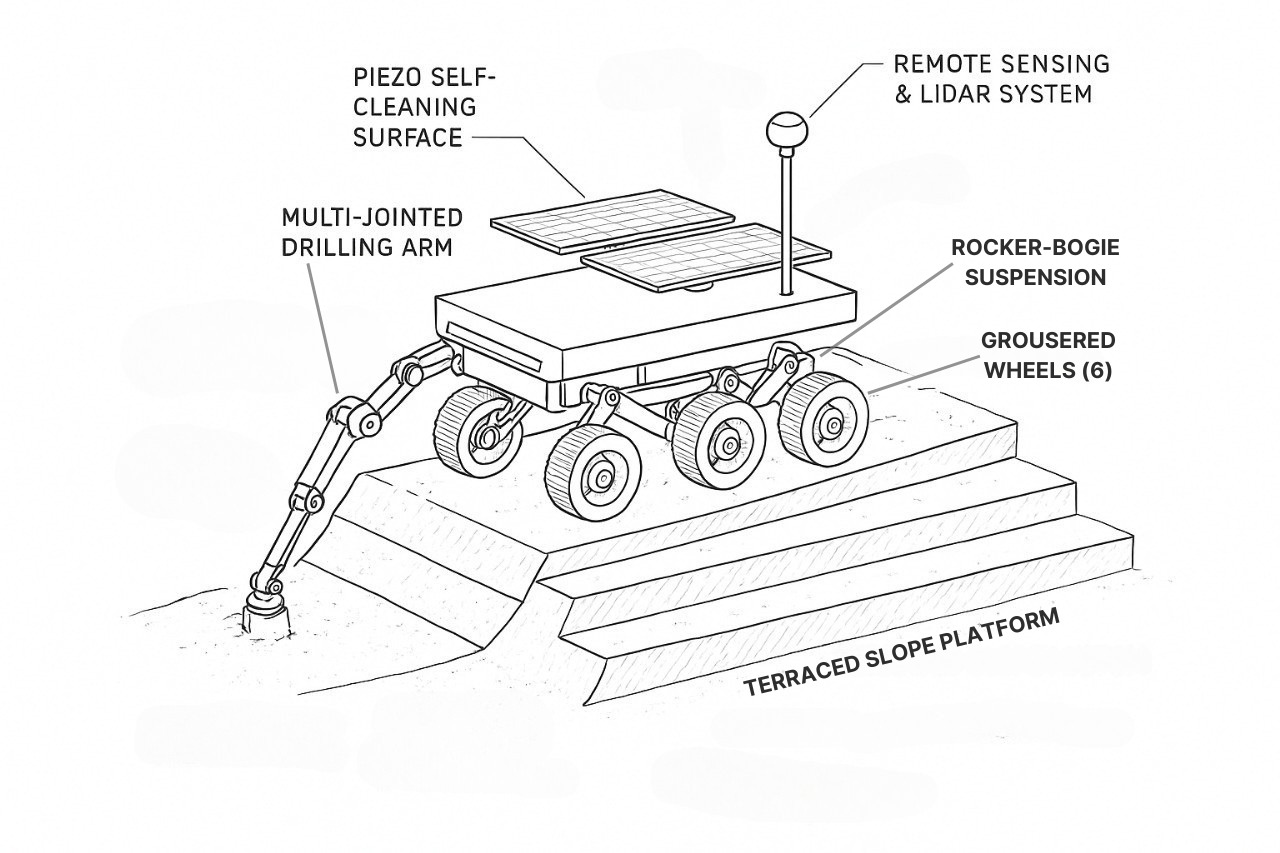}
    \caption{Schematic diagram of rover highlighting important functionalities}
    \label{fig:Rover}
\end{figure}

The schematic diagram of the various functions of rover is shown in figure \ref{fig:Schematic}. A preliminary design of the rover with the important parts is shown in figure \ref{fig:Rover}.

\subsection{Derivation of Power and Energy Requirements for the Rover}

The total force a rover must overcome while traversing Martian terrain includes:

\begin{itemize}
    \item Rolling resistance: \( F_{\text{roll}} = \mu_{rr} \cdot m \cdot g \cdot \cos(\theta) \)
    \item Aerodynamic drag: \( F_{\text{drag}} = \frac{1}{2} \cdot \rho \cdot A \cdot v^2 \cdot C_d \)
    \item Gravitational component due to slope: \( F_{\text{slope}} = m \cdot g \cdot \sin(\theta) \)
\end{itemize}

We observe that all components of the total force act in the same direction, i.e. against the motion. Thus, the total force required to move the rover (in the direction of the motion) is:
\[
F_{\text{total}} = \mu_{rr} \cdot m \cdot g \cdot \cos(\theta) + \frac{1}{2} \cdot \rho \cdot A \cdot v^2 \cdot C_d + m \cdot g \cdot \sin(\theta)
\]

The energy required is the dot product of the force vector and the displacement vector:
\[
E = \vec{F}_{\text{total}} \cdot \vec{D}
\]

Since the force required and displacement act in the same direction, the angle between the vectors is 0. Thus, we substitute the total force:
\[
E = D \cdot \left[ \mu_{rr} \cdot m \cdot g \cdot \cos(\theta) + \frac{1}{2} \cdot \rho \cdot A \cdot v^2 \cdot C_d + m \cdot g \cdot \sin(\theta) \right]
\]

This expression gives the total energy (in watt-seconds or joules) required for the rover to traverse a distance \( D \) at constant velocity \( v \).

\begin{table}[H]
\centering
\renewcommand{\arraystretch}{1.3}
\begin{tabular}{|p{4.2cm}|p{3.5cm}|p{6.2cm}|}
\hline
\textbf{Parameter} & \textbf{Value} & \textbf{Notes} \\
\hline
Gravitational acceleration on Mars (g) & $3.721$ m/s\textsuperscript{2} & $38\%$ of Earth's gravity   \\
Rolling resistance coefficient ($\mu$\textsubscript{rr}) & $\sim0.1$ & Estimated for Volcanic Terrain \cite{langley2003mars_tumbleweed} \\
Atmospheric density on Mars ($\rho$) & $0.020$ kg/m\textsuperscript{3} & \\
Frontal area of rover (A) & $5.5-6.5$ m\textsuperscript{2} & Through Aspect Ratio approx. $2.5:2.5$ (e.g., Perseverance approx. $5.94$ m\textsuperscript{2}) \\
Drag coefficient (C\textsubscript{d}) & $0.5$ & Estimated for Geometry of Designed Rover\cite{langley2003mars_tumbleweed} \\
\hline
\end{tabular}
\caption{Constants for calculating power and energy requirements of Mars rovers.}
\label{tab:mars_constants}
\end{table}

Noting the constants in Table 2, we substitute the values in our equation for total energy for rovers moving at a constant velocity of 0.01 m/s, and find the simplified equation as follows:
\[
E = D \cdot \left( 0.3721 \cdot m \cdot \cos(\theta) + 0.00000045 + 3.721 \cdot m \cdot \sin(\theta) \right)
\]

The horizontal distances from the proposed base of operations to the summit resource locations were derived from MOLA-based planning plots. For analysis, we have picked one volcano from each of the geographical groups highlighted in table \ref{tab:volcano_mola_tiles} (Olympus Mons from first group, Arsia Mons from group 2 and Elysium Mons from group 3). The resulting one-way distances are as follows for these three volcanoes:

\begin{itemize}
  \item Olympus Mons: \(\mathbf{231.5\ \text{km}}\)  
  \item Arsia Mons: \(\mathbf{324.1\ \text{km}}\)  
  \item Elysium Mons: \(\mathbf{509.3\ \text{km}}\)
\end{itemize}

An average traverse speed of 0.05\(\mathrm{m/s}\) was assumed for the energy estimate in this analysis based on previous missions.\cite{wilcox2011athlete}

The following table summarizes the computed traverse energy for each volcano where all parameters are as defined previously.

\begin{table}[H]
\centering
\label{tab:traverse_energy}
\renewcommand{\arraystretch}{1.2}
\begin{tabular}{|l|r|r|}
\hline
\textbf{Target Resource} & \textbf{Distance (km)} & \textbf{Traverse Energy (kWh)} \\
\hline
Olympus Mons & 231.5 & 96.1 \\
Arsia Mons   & 324.1 & 108.5 \\
Elysium Mons & 509.3 & 145.6 \\
\hline
\end{tabular}
\caption{One-way traverse energy estimates (mechanical energy) at \(v = 0.05\ \mathrm{m/s}\).}
\end{table}
The supplementary energy requirements for operational purposes on Mars extend beyond simple locomotion, encompassing various scientific and mission-critical tasks such as drilling, sampling, communication, and data transmission. Drilling tools, for example, demand significant energy depending on the depth and substrate. The Rock Abrasion Tool (RAT), used on NASA’s Spirit and Opportunity rovers, operates at an average power consumption of 30 watts.\cite{nssdc2003mars} More advanced systems like the Icebreaker-3 drill, designed for deep subsurface exploration, require between 200 watts of power and energy depending on drilling duration and depth\cite{nasa_20160006480}. Meanwhile, low-mass coring devices such as the robotic MASA drill maintain power efficiency with sub-100 watt operation but may still consume 40–400 watt-hours depending on operational cycles \cite{anttila2005}. These drilling and sampling systems are complemented by robotic sample handling mechanisms, such as the complex caching system aboard NASA's Perseverance rover, which, while not directly quantified in terms of energy consumption, likely contributes a non-trivial load due to its intricate actuators and control electronics.\cite{jpl2020_perseverance_sample_system} Additionally, communication systems like those onboard Curiosity—which utilize both X-band and UHF links—rely on a steady 110-watt output from the Multi-Mission Radioisotope Thermoelectric Generator (MMRTG)\cite{nasa_mars_mmrtg} to support data relay and system health checks. Taken together, these operations demonstrate that energy planning for Mars missions must account for diverse and often high-demand subsystems beyond mobility alone.

\begin{table}[H]
\centering
\begin{tabular}{|p{3.5cm}|p{3cm}|p{4cm}|p{5cm}|}
\hline
\textbf{Operation} & \textbf{Average Power Consumption (W)} & \textbf{Total Energy Consumption (Wh)} & \textbf{Remarks}\\
\hline
Drilling & 30 watts & 45-60 watt-hours & Rock Abrasion Tool (RAT)\\
 & 200 watts & 16.67 - 33.33 watt-hours &   Icebreaker-3\\
Low-Mass Coring Devices & $<$100 watts & Varying (40-400 watt-hours) & Robotic ``MASA" Drill\\
Communications & $110$ watts & Varying & MMRTG (Curiosity Rover)\\
\hline
\end{tabular}
\caption{Estimated power consumption of various Mars sample handling and drilling operations.}
\label{tab:power_consumption}
\end{table}

\section{Discussions and Conclusions}

The purpose of this work is to investigate the feasibility of resource extraction in Martian volcanic terrains. We have used six volcanoes of Mars as case studies to evaluate the setting up of a resource extraction programme in Martian volcanic terrains. Volcanic areas on Mars contain abundant resources like subsurface water ice and metals such as iron, nickel, and titanium. However, the extraction of these resources also presents various operational risks. Steep slopes, regolith instability, and pervasive dust require that extraction strategies be tailored to the terrain rather than relying on generic ISRU models. To address these concerns, we first performed a slope-stability analysis for all three volcanoes outlined above. Additionally, for Olympus Mons, we also utilized TES data to further understand subtle geological complexities of volcanic terrain. Based on our assessment of the complexity of Martian volcanic terrain, we arrived at the conclusion that slope-stabilization measures such as terracing, anchored drilling and implementation of terrain adaptive robotic systems can make mining systems viable in Martian volcanic regions by providing a practical balance between safety and accessibility. In this work, we have utilized MOLA dataset for slope-stability analysis. Other crucial data like Compact Reconnaissance Imaging Spectrometer for Mars (CRISM)\cite{crism_pds} hyperspectral cubes, High Resolution Imaging Science Experiment (HiRISE)/Context Camera (CTX) imagery\cite{hirise_uofa} can be utilized in future follow-up works so that surface mineralogy, grain size, albedo, and slope variations of Martian volcanic terrains can be determined with utmost precesion. 

We also performed a comparative analysis between roads and rails as transport infrastructure. The comparative analysis highlights a clear trade-off. Although roads constructed with compacted regolith or sulfur concrete can be implemented quickly and flexibly to meet early mission requirements but their surfaces degrade rapidly under Martian dust and high-frequency rover traffic. Rails, by contrast, require greater upfront planning and material investment but deliver higher efficiency, reduced maintenance, and resilience against dust accumulation in the long term. Henceforth, we propose that a multi-phased approach appears most viable for setting up the Martian volcanic mining programme: roads for initial exploration and settlement, followed by rail-based infrastructure for sustainable, scaled-up extraction operations.
This multi-phase layered strategy not only provides a feasable way to initiate and sustain mining operations in Mars but in a larger perspective also aligns with broader discussions on the logistical challenges of Mars colonization. 

The slope-mining rover proposed here (approx. 2.1 t) shows an effort to reconcile terrain adaptability with efficiency. Features such as rocker–bogie suspension, anchoring-enabled drilling arms, and dust-mitigation systems provide the necessary robustness for high-slope volcanic flanks. Critically, autonomous navigation tools enable the rover to operate with minimal Earth-based intervention, an essential feature given communication delays. Such autonomy ensures that rovers can adapt to dynamic surface conditions, including high-slope navigation in real time, which strengthens mission resilience and extends operational lifespans. We also did energy modelling of the rover to estimate power requirements of sustained operations. Energy modeling reveals that mobility, drilling, dust removal, and communication impose competing demands on limited power budgets. Hybrid energy solutions that combine solar arrays with compact nuclear sources appear best suited for supporting extended operations. Operational efficiency of rovers will depend heavily on route optimization and scheduling to align rover activity with environmental windows.

In conclusion, this study offers an integrated framework for advancing resource extraction on Mars by addressing the unique mechanical and environmental challenges of Mars, including reduced gravity, abrasive basaltic regolith, and pervasive dust deposition. Roads offer immediacy, rails provide long-term efficiency, and rover systems must balance stability, adaptability, and energy management. The common thread across all findings is adaptability—whether in engineering design, mission sequencing, or energy use. These engineering interventions echo the broader lesson that resource use has always underpinned human development and must now evolve for extraterrestrial contexts. Future work should focus on coupled system simulations that integrate geological datasets with rover performance models and on analog field trials in terrestrial volcanic regions. Such developments will help bridge the gap between theoretical feasibility and practical readiness, bringing humanity closer to establishing a sustainable foothold on Mars. 

Overall, Martian resource extraction missions will require synergistic engineering solutions that integrates mobility, infrastructure resilience, environmental mitigation, and autonomous operations to ensure sustained access to valuable resources under the planet’s extreme conditions.

\section*{Acknowledgments}

We would like to express our sincere gratitude to \textbf{Dr. Nishchal Dwivedi}, Senior Director, Pangea Society, for his invaluable insights and guidance throughout the course of this research. His expertise and constructive feedback played a crucial role in shaping the direction and depth of this work.

\printbibliography
\end{document}